 \documentclass[aps,prl,twocolumn,superscriptaddress,showpacs]{revtex4}
\usepackage{epsfig}

 \usepackage{dcolumn}
 \usepackage{bm}

\begin{document}

\title{Magnetic field induced rotation of the d-vector in 
the spin triplet superconductor Sr$_2$RuO$_4$}

 \author{James F. Annett}
 \affiliation{H. H. Wills Physics Laboratory, University of Bristol,  
Tyndall Ave, BS8-1TL,
UK.}

 \author{B. L. Gy\"orffy}
 \affiliation{H. H. Wills Physics Laboratory, University of Bristol, 
Tyndall Ave, BS8-1TL,
UK.}

 \author{G. Litak}
 \affiliation{Department of Mechanics, Technical University of Lublin,
Nadbystrzycka 36, PL-20-618 Lublin, Poland.}

 \author{K. I. Wysoki\'nski}
 \affiliation{Institute of Physics, M. Curie-Sk\l{}odowska University,
Radziszewskiego 10, PL-20-031 Lublin, Poland}
\affiliation{ Max Planck Institut f\"ur 
             Physik komplexer Systeme,  
             D-01187 Dresden, Germany} 
 \date{ \today}

\begin{abstract}
In zero magnetic field the superconductor Sr$_2$RuO$_4$ is believed to 
have a chiral spin triplet pairing state in which the 
gap function d-vector is aligned along the crystal $c$-axis.  
Using a phenomenological but orbital
specific  description of the  spin dependent electron-electron 
attraction and a realistic
quantitative account of the electronic structure in the normal state we
analyze the orientation of the spin triplet Cooper pair  d-vector in response to 
an external $c$-axis magnetic field. 
We show that for suitable values of the model 
parameters a c-axis field of only $20$ mT is able 
to cause a reorientation phase transition of the 
d-vector from along $c$ to the $a-b$ plane,
in agreement with recent experiments.
\end{abstract}

\pacs{PACS numbers:
             74.70.Pq,                 74.20.Rp,        74.25.Bt     }

\maketitle

\section{Introduction}

The $1.5$ K superconductor Sr$_2$RuO$_4$  \cite{maeno1994} 
is widely believed to be a rare example of a spin-triplet
Cooper paired system \cite{maeno2001,mackenzie2003}.  Unlike
heavy fermion materials which are possible candidates
for spin triplet pairing (such as UPt$_3$) there are no $4f$ or $5f$ heavy elements 
in the unit cell, and so the effects of spin-orbit coupling should be 
relatively weak.  In this respect paring in
Sr$_2$RuO$_4$ should be closely analogous to the case of superfluid $^3$He.
One should expect that
weak or moderate $B$-fields would be able to rotate the d-vector
order parameter and induce domain walls, textures and other topological defects.
 As in the case of $^3He$, the experimental observation of such d-vector rotations
would be a decisive test of the pairing symmetry.  Systematic study of
such rotations as a function of various physical parameters would both 
confirm the pairing symmetry and also place strong constraints on the
pairing mechanism and microscopic Hamiltonian parameters.

One of the strongest pieces of evidence for spin triplet
Cooper pairing in Sr$_2$RuO$_4$ was the observation that the
electronic spin susceptibility $\chi_s(T)$ remained constant below $T_c$
for magnetic fields in the a-b plane \cite{ishida1998,duffy2000}.
This observation is inconsistent with a spin-singlet
pairing state ($s$ or $d$-wave) for which $\chi_s(T) = \chi_n Y(T)$,
where $\chi_n$ is the normal state Pauli spin susceptibility and
$Y(T)$ is the Yoshida function.  On the other hand
the observations  would be immediately consistent with a chiral symmetry
paring state of the form 
\begin{equation}
              {\bf d}({\bf k}) = (\sin{k_x}+i\sin{k_y})\hat{\bf e}_z,
          \label{eq:chiral-a}
\end{equation}
which below we shall refer to  as pairing state $(a)$.
For such a pairing state 
is is well known \cite{leggett1975} that the spin susceptibility
has the following tensor form
\begin{equation}
  \hat{\chi}_s(T) =  \chi_n \left( \begin{array}{ccc} 1 & 0 & 0 \\
  0 & 1 & 0 \\
  0 & 0 & Y(T)  \end{array}\right). \label{eq:chi-a}
\end{equation}
and so the susceptibility tensor is constant as a function of $T$ 
for field directions
perpendicular to the ${\bf d}$ vector.  The $^{17}O$ Knight shift
experiments performed in (ab) plane magnetic fields \cite{ishida1998}
and neutron scattering  experiments \cite{duffy2000} both observed
a constant susceptibility below $T_c$, a result which \emph{uniquely}
points to a chiral triplet paring state with ${\bf d} \parallel  \hat{\bf 
e}_z$.

More recently Murakawa {\it et al.} \cite{murakawa2004} have measured 
the $^{101}$Ru-Knight shift of 
 Sr$_2$RuO$_4$ in a superconducting state under the influence
of magnetic field  parallel to the c-axis.   In contradiction
to expectations from Eqs. (\ref{eq:chiral-a}) and (\ref{eq:chi-a})  they found 
that its value is \emph{also 
unchanged from the normal state value below} $T_c$. 
They remarked that this result would be consistent with 
Eq. (\ref{eq:chi-a})  only if it is assumed that the ${\bf d}$ vector
rotates away from the ${\bf \hat{e}_z}$ direction under the influence of
the external field. Such a rotation of the ${\bf d}$ vector is well
known in the case of superfluid $^3$He-A \cite{leggett1975}, where
to minimize free energy ${\bf d}$ orients itself 
perpendicular to the external field unless
pinned by surface effects.
However, in  Sr$_2$RuO$_4$ it is expected that spin-orbit coupling would
fix the chiral state ${\bf d}$ vector to the 
crystal c-axis \cite{ng2000a,ng2000b,ng2000c,annett2006}. 
Murakawa {\it et al.} \cite{murakawa2004} were able to measure $\chi_s(T)$ down to
fields as low as $20$ mT at $80$ mK, showing that it was 
equal to $\chi_n$ to within experimental accuracy at all
temperatures and fields measured below $B_{c2}$ ($75$ mT for c-axis fields).
The implication of this result, that ${\bf d}$ vector 
rotation must occur at below $20$ mT,
therefore places very strong constraints on the
strength of pinning by the spin-orbit coupling.

In this work we examine the combined effects of spin-orbit coupling
and c-axis magnetic field on the chiral state of Sr$_2$RuO$_4$.  
We show that a realistic physical model, which we have previously shown
to be consistent with a wide range of experimental data, allows us to find
reasonable parameter ranges where the chiral state 
Eq.~(\ref{eq:chiral-a}) is stable
in zero field, but where a transition to another state occurs even for 
fields of order $20$ mT or smaller. The minimum energy states
in finite c-axis field are not simply the chiral state
with ${\bf d}$ vector rotated to the $a-b$ plane, 
but rather are non-chiral pairing states
of the form
\begin{eqnarray}
  {\bf d}({\bf k}) &=& (\sin{k_x},\sin{k_y},0)  \label{eq:deltab}  \\
   {\bf d}({\bf k}) &=& (\sin{k_y},-\sin{k_x},0)  \nonumber
\end{eqnarray} 
on the $\gamma$ Fermi surface sheet (which below we refer 
to as $(b)$ and $(c)$ respectively).
The spin susceptibility for either of these states is
of the form \cite{leggett1975,annett1990}
\begin{equation}
\hat{\chi}_s(T) =  \frac{1}{2} \chi_n \left( \begin{array}{ccc} 1+Y(T) & 0 & 0 \\
  0 & 1+Y(T) & 0 \\
  0 & 0 & 2  \end{array}\right), \label{eq:chi-bc}
\end{equation}
corresponding to constant spin susceptibility for c-axis fields  
consistent with the results of Murakawa {\it et al.} \cite{murakawa2004}.

Since these states are symmetry distinct from the chiral state 
Eq.~(\ref{eq:chiral-a}) the transition from one to another  
should occur at a finite external field, $B_t$, 
and it should be accompanied by a finite entropy change $\Delta S$. Below we map
out a generic phase diagram for the transitions between pairing states 
of Sr$_2$RuO$_4$
in a c-axis field, assuming parameter values consistent with the
Murakawa's {\it et al.} experiments \cite{murakawa2004}.  
 We also estimate the entropy 
change associated
with the transition, and comment on  its experimental observability.
Our preliminary results have been presented in \cite{annett2007}.

\section{The model}

As the pairing mechanism operating in the strontium ruthenate  
superconductor is not known \cite{maeno2001,mackenzie2003}
it is   important to understand the experiments on the basis
of semi-phenomenological models \cite{agterberg1997,zhitomirsky2001}.
Specifically we write the
effective pairing Hamiltonian as
 \begin{eqnarray}
 \hat{H}& =&  \sum_{ijmm^\prime,\sigma}  \left( ( \varepsilon_m -
 \mu) \delta_{ij} \delta_{mm^{ \prime}} - t_{mm^{ \prime}}(ij)  \right)  \hat{c}%
^+_{im \sigma} \hat{c}_{jm^{ \prime} \sigma}   \nonumber  \\
 & & + \mathrm{i}  \frac{ \lambda}{2}  \sum_{i, \sigma \sigma^{ \prime}\kappa }
 \sum_{mm^{ \prime}}  \varepsilon^{ \kappa mm^{ \prime}}
 \sigma^{ \kappa}_{ \sigma \sigma^{ \prime}} c^+_{i m \sigma} c_{i
m^{ \prime} \sigma^{ \prime}} \nonumber \\
&& -  \frac{1}{2}  \sum_{ijmm^{ \prime}}  \sum_{\alpha\beta\gamma\delta}
U_{mm^{ \prime}}^{ \alpha \beta, \gamma \delta}(ij) 
 \hat{c}^+_{im \alpha } \hat{c}^+_{jm^ \prime \beta } \hat{c}_{jm^ \prime \gamma } 
 \hat{c}_{im \delta },
 \label{hubbard}
 \end{eqnarray}
where $m$ and $m^{ \prime}$ refer to the three Ru $t_{2g}$ orbitals 
$a=d_{xz}$, $b = d_{yz}$ and $c = d_{xy}$, and $i$ and $j$ label the sites of a body
centered tetragonal lattice. The hopping integrals $t_{mm^{ \prime}}(ij)$ and
site energies $ \varepsilon_m $ were fitted to reproduce the experimentally
determined Fermi surface  \cite{mackenzie1996,bergemann2000}. 
$\lambda$ is the effective 
Ru 4d spin-orbit coupling parameter, and 
the effective Hubbard  parameters 
$U_{mm^{ \prime}}^{ \alpha \beta, \gamma \delta}(  ij)$  are 
generally both spin \cite{baskaran1996,spalek2001} 
and orbital dependent \cite{agterberg1997,zhitomirsky2001,annett2002}.

Our approach is based on self consistent solution of the Bogolubov-deGennes
equations for the model (\ref{hubbard}) in the spin triplet channel for each possible 
symmetry distinct order parameter.
We have shown elsewhere \cite{annett2002,annett2003} that a minimal realistic model
requires two Hubbard $U$ 
parameters: $U_{\parallel}$ for 
nearest neighbor in-plane interactions between Ru $d_{xy}$ orbitals (corresponding to the $\gamma$
Fermi surface sheet) and $U_{\perp}$ for out of plane nearest neighbor interactions between
Ru $d_{xz}$ and $d_{yz}$  orbitals (corresponding to the $\alpha$ and $\beta$ Fermi surface sheets).
When these two parameters are chosen to be spin independent constants (for $\lambda=0$) 
and to give the experimental $T_c=1.5$K on all three $\alpha$, $\beta$ and $\gamma$ 
Fermi surface sheets, then the calculated specific 
heat, thermal conductivity  and $a-b$ plane  penetration depth
are in very good agreement with  
experiments \cite{nishizaki2000,tanatar2001,bonlade2000}. 
  
\begin{figure}[tbp]
\epsfig{file=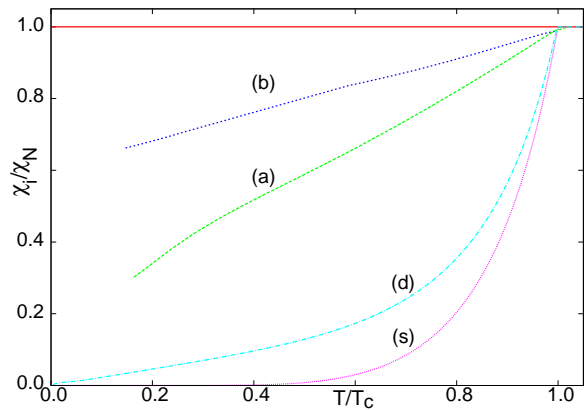,width=5.5cm,angle=-90}
 
\caption{ (Color online) Temperature dependence of the normalized 
spin susceptibility in the superconducting state for a number of 
order parameters.  The curves marked (a) and (b) show the c-axis susceptibility for Sr$_2$RuO$_4$
and chiral symmetry state, $(a)$, defined by Eq.~\ref{eq.delta2} and Eq.~\ref{eq.delta3},  
and the ab-plane susceptibility of the non-chiral triplet state, of symmetry $(b)$, 
corresponding to Eq.~\ref{eq:deltab}.  
Curves labelled (s) and (d) show the Yoshida function for single band superconductor with two dimensional
tight binding spectrum and order parameter of $s$ and $d$ wave symmetry, respectively.
\label{fig:chi}}
\end{figure}

Figure (\ref{fig:chi}) shows the susceptibilities 
calculated within the model for the (a) and (b) states 
with the symmetries given by Eqs.~\ref{eq:chiral-a} and \ref{eq:deltab}. 
For the chiral state (a) the 
triplet d-vector is defined 
in all 3 bands as
${\bf d}({\bf k}) = \Delta_{mm^{\prime }}({\bf k})\hat{\bf e}_z$,
 with   $\Delta_{mm^{\prime }}({\bf k})$ denoting contributions from 
different orbitals which are given by
\begin{equation}
\Delta _{cc}({\bf k})=\Delta _{cc}(T)(\sin {k_{x}}+i\sin{k_{y}})
\label{eq.delta2}
\end{equation}
for the Ru $c (=d_{xy})$ orbitals and,
\begin{equation}
\Delta _{mm^{\prime}}({\bf k})= 
\Delta _{mm^{\prime}} 
(\sin {\frac{k_{x}}{2}}\cos {\frac{k_{y}}{2}}+i \sin {\frac{k_{y}}{2}}\cos {\frac{k_{y}}{2}}
)\cos {\frac{k_{z}c}{2}}  
\label{eq.delta3}
\end{equation}%
for $m,m^{\prime }=a,b$ corresponding to the Ru $d_{xz}$ and $d_{yz}$
 orbitals  \cite{annett2002}. 
The susceptibility results, for the chiral (a) state show that the 
 dependence of $\chi$ for c-axis B-fields 
is is relatively structureless. The relevant Yoshida function $Y(T)$ 
is equal to unity at $T=T_c$ and drops essentially linearly to zero at 
$T=0$, consistent with the expectation from Eq.{\ref{eq:chi-a}}
 and the existence of the horizontal line node in the gap on
  $\alpha$ and $\beta$ \cite{litak2004}.
In contrast, we show in Fig.~\ref{fig:chi} that for the 
non-chiral (b) pairing state the susceptibility for 
a-axis B-fields decreases approximately linearly towards $\chi_n/2$ at $T=0$, 
again consistent with the expectations of Eq.~\ref{eq:chi-bc}. 
 These results confirm that in spite of very complicated 
 character of the order parameter (see ref. \cite{litak2004}) 
 having zeros on the $\alpha$ and $\beta$ sheets of the
Fermi surface, the overall behavior of the calculated susceptibility 
tensor is consistent with the expectations from
superfluid $^3 He$ \cite{leggett1975,annett1990}.

In zero magnetic field, but in the presence of non-zero spin-orbit interaction $\lambda$ 
the model predicts that the ground state is the chiral state 
$(a)$ with d-vector along the
crystal c - axis, provided that the 
spin-orbit coupling leads to a small spin dependence of the effective pairing
interaction \cite{annett2006}.  Choosing $U'\equiv U^{\uparrow\downarrow}$
 about  1\% larger than $U \equiv U^{\uparrow\uparrow}=U^{\downarrow\downarrow}$
 is sufficient to stabilize the chiral state even for
large spin orbit coupling. In contrast, for spin-independent interactions, $U'=U$,
the alternative pairing states $(b)$ and $(c)$  ($(d)$ and $(e)$ \cite{annett2006}) are the ground 
states for any value of $\lambda<0$ ($\lambda>0$) \cite{sigrist2005}.

\section{Phase diagram in a magnetic field}

\begin{figure}[tbp]
\epsfig{file=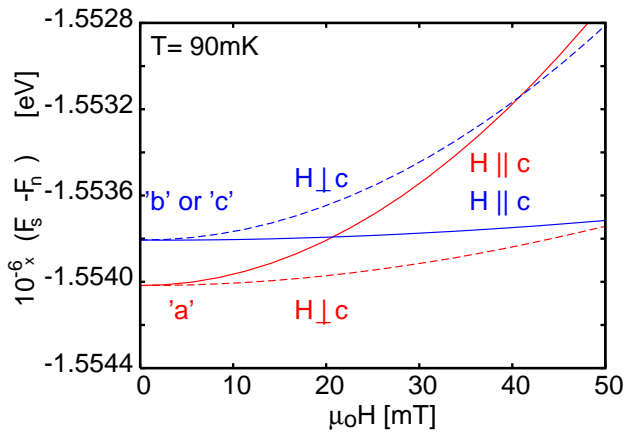,width=6.5cm,angle=-90}


\caption{(Color online) Condensation free energy at $T=90$mK for three triplet 
order parameters of different symmetries, $(a)$, $(b)$ and $(c)$ 
as a function of
  external field $H\parallel c$ (solid lines) and $H\perp c$ 
(dashed lines), respectively. 
The parameters used for calculations are: $\lambda=-0.02t$ where $t=80$meV is the effective 
$\gamma$ band nearest neighbor hopping integral, and $U'=1.0011U$.
\label{fig2}}
\end{figure}

In Fig.~\ref{fig2} we show the free energies of the $(a)$ $(b)$ and $(c)$
symmetry pairing states as a function of external magnetic field both for
$H\parallel c$ and $H\perp c$.  The model parameters were chosen to make the 
chiral state $(a)$ stable at zero field. For the field in ab-plane the
free energy of the chiral state $(a)$ increases with field more slowly 
than for the $(b)$ and $(c)$ states. This means that in large  in-plane fields
 the $(a)$ state becomes 
 relatively more stable than $(b)$ or $(c)$. 
On the other hand,  in an c-axis field the chiral phase increases 
its free energy  much faster than the
other phases, until
at a certain critical field, $B_t$, the $(b)$ or $(c)$ solutions 
become more stable.
 Therefore at the field $B_t$ we expect a ``spin flop'' type phase transition from a d-vector 
oriented along the $c$-axis
to one where the d-vector lies in the $a-b$ plane.   For the parameter values used in
Fig.~\ref{fig2} this critical field is $B_t \approx 20$ mT.
to one where the d-vector lies in the $a-b$ plane.   For the parameter values used in

This prediction is consistent with what one expects
from the analogous superfluid state in $^3He$-A.  
In bulk $^3He$ there is no preferred symmetry direction in space
and so the d-vector simply rotates continuously to 
remain perpendicular to the applied field. 
But in thin  films the d-vector is pinned, and only 
rotates at a finite critical field, 
known as the Freedericksz
transition \cite{vollhardt1990}.  In the case of 
Sr$_2$RuO$_4$ the transition is not simply a rotation
of the chiral d-vector but is also a transition 
from a chiral to non-chiral pairing state of different symmetry. 
The two distinct solutions shown in Fig.~\ref{fig2} 
(note that $(b)$ and $(c)$ are essentially degenerate)
have different entropies  and 
hence this Freedericksz-like spin-flop  is 
a first order thermodynamic phase transition.

\begin{figure}[tbp]
\epsfig{file=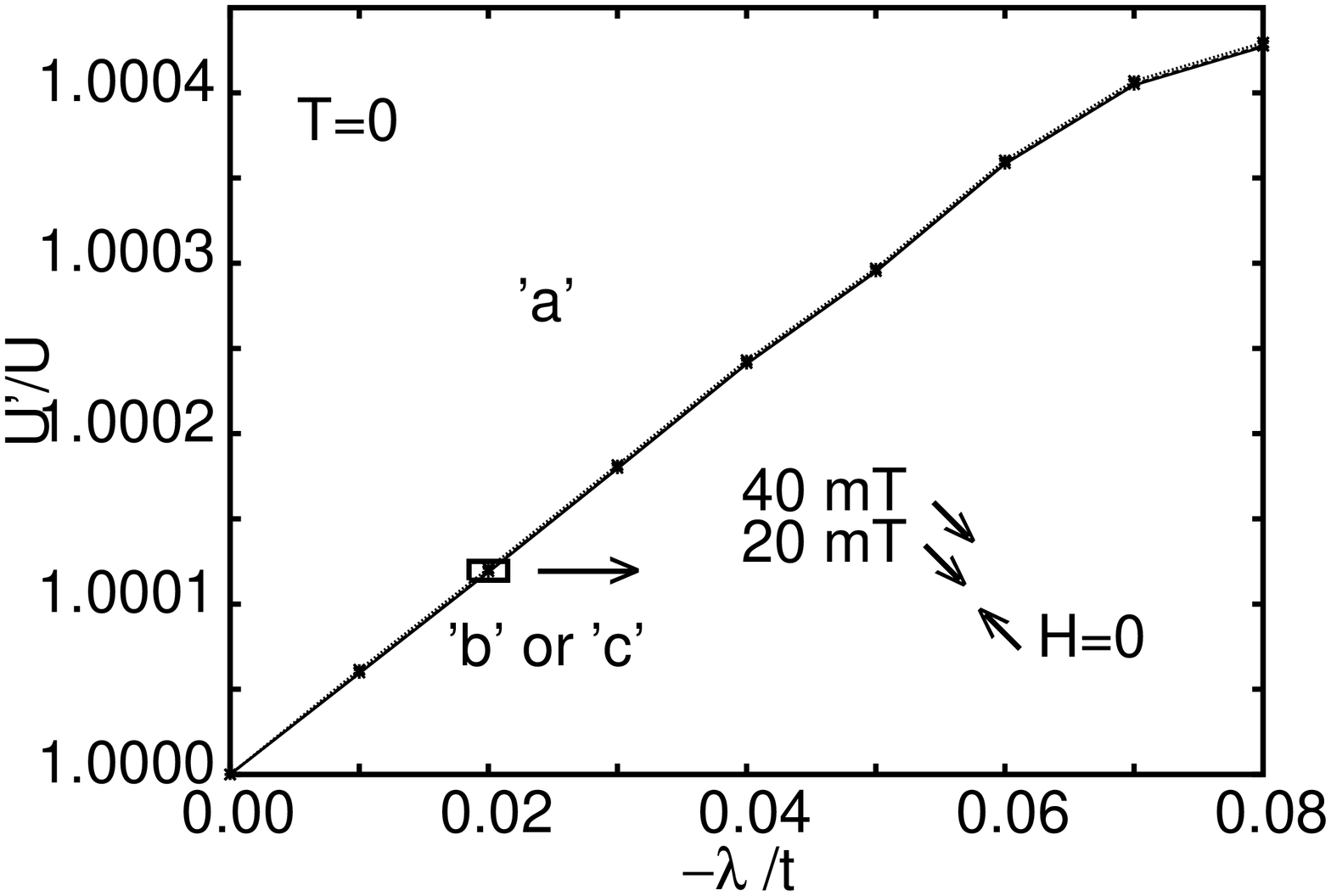,width=6.5cm,angle=-90}
\vspace{-3.7cm}

\hspace{3.5cm}
\epsfig{file=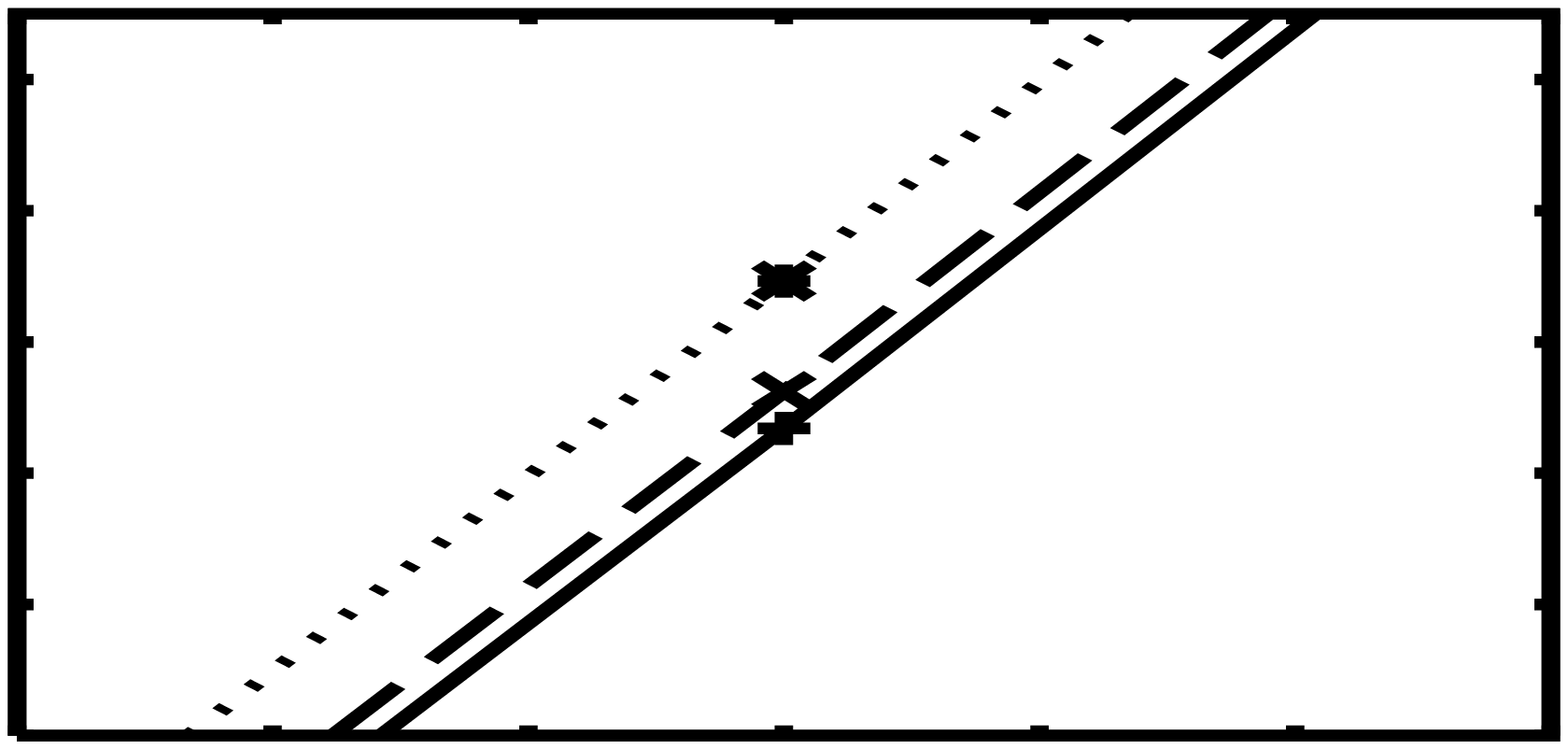,width=3.5cm,angle=-90}
\vspace{0.50cm}

\caption{Critical interaction $U'/U$ as a function of $\lambda$ 
in the presence of external field $H\parallel c$ at $T=0$. The 
critical c-axis field of the $(a)$ to $(b)$ or $(c)$ spin flop 
transition depends on how close the physical parameters lie  to 
the zero-field $(a)-(b)/(c)$ phase boundary.
\label{fig3}}
\end{figure}

 One can also ask whether the specific case of 
Fig~\ref{fig2} is typical for a general set of model parameters. 
 In our effective Hubbard model the spin-orbit 
interaction enters the Hamiltonian both directly
 {\it via} $\lambda$ in Eq.~(\ref{hubbard}), and also in 
the spin dependent pairing potential $U'/U$. These are not really
 independent, since a full theory, such as a spin 
fluctuation model \cite{eremin2002,monthoux2005,nomura2008}, would 
include both effects 
 on the same footing.  In Fig.~\ref{fig3} we show 
a part of the full phase diagram of our model \cite{annett2006} 
 for a specific choice of the parameters $\lambda$ and $U'/U$.  The 
inset shows the effect of the c-axis magnetic field in 
shifting the phase boundary  between the chiral state 
$(a)$  and the alternative $(b)$ and $(c)$ states. At any point
in the phase diagram where $(a)$ is stable for zero field, 
there is a definite critical field for which the
spin flop to $(b)$ or $(c)$ takes place.  However it is 
only a small region of the phase diagram, close to the 
$H=0$ phase boundary, where the transition takes place 
in fields as small as $20-40$ mT. The implications 
of the Murakawa {\it et al.} experiment \cite{murakawa2004} 
are that the parameters do indeed lie in this region.  The 
alternative, is that
the spin flop only takes place at a much larger field, and 
thus would never be observed in any c-axis 
field below $H_{c2}=75$ mT, in contradiction to the 
simplest interpretation of Murakawa {\it et al.}'s result.

\begin{figure}[tbp]
\epsfig{file=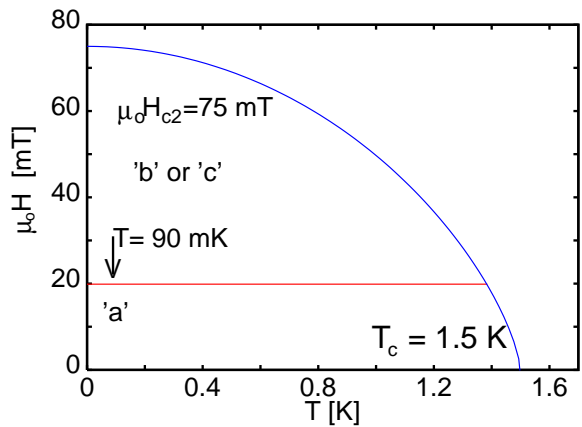,width=6.5cm,angle=-90}
\caption{(Color online) The expected phase diagram in presence of external field $H
\parallel c$ in the region below $H_{c2}$. 
\label{fig4}}
\end{figure}

If the relevant parameters do allow the d-vector rotation 
to occur in a relatively low field, then we expect  a phase diagram similar to Fig.~\ref{fig4}.
This 
hypothesis makes a definite and clear testable prediction, namely
the existence of a new first order phase boundary 
in the $H-T$ plane, which has not yet been observed.
 A phase diagram  of the type
shown in Fig.~\ref{fig4}  is required 
from the experiments \cite{murakawa2004,xia2006}.
The NQR experiments \cite{murakawa2004} 
indicate the state with d-vector lying  in the (ab) plane is
realized in a c-axis magnetic field of order $B=20$ mT and larger, and that
this d-vector orientation must be present at these fields for
 all
temperatures below $T_c$.  
On the other hand the nonzero Kerr signal obtained at fields 
around $10$ mT \cite{xia2006} (see the points in Fig. (3a) of this reference
and the corresponding discussion in the text) 
is a clear indication  that at these fields the chiral
state is stable. Accepting the usual interpretations of these two experiments
one is forced to accept the general structure of the phase diagram 
as in Fig.~\ref{fig4} as the simplest picture which is consistent with both experiments,  with a possible  
margin of uncertainty of about $10$ mT  in the position of the $B=B_t$ transition line. 

\begin{figure}[tbp]

\epsfig{file=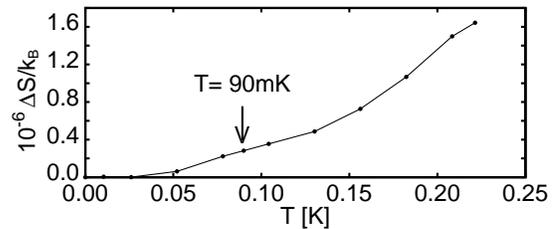,width=3.5cm,angle=-90}
\caption{The temperature dependence of the  entropy 
jump along the transition line $B=B_t=20$ mT and  $\lambda=-0.02t$. 
\label{fig5}}
\vspace*{-0.5cm}
\end{figure}

An obvious question is why has the predicted phase transition in Fig.~\ref{fig4} not been observed
experimentally?  Experimental data \cite{mao2000} show
multiple superconducting phases at the magnetic field  parallel
to the (ab) plane, but none have been reported for fields along the c-axis. 
 The answer to this question we suggest is that the phases 
in question have essentially identical values of 
the quasi-particle energy gap,
$|{\bf d}({\bf k})|$ on the Fermi surface,  
and therefore to a very high level of accuracy they
have essentially identical entropy.  This is because 
to leading order only the direction of 
${\bf d}({\bf k})$ changes at each point on the Fermi 
surface, and not its magnitude. 
To confirm this we have calculated the entropy change 
along the transition line
$B=20$ mT  and this is shown in the Fig.~\ref{fig5}.  
The total change in entropy is very small, of order
$10^{-6} k_B$ per formula unit, suggesting that it might not
 have been detected in specific heat experiments. 
The entropy change in Fig.~\ref{fig5} corresponds to 
a latent heat of at most $0.003$mJ/mol, compared to the
zero field specific heat jump $\Delta C/T$ of about 
$28$mJ/K$^2$mol at $T_c=1.5$K \cite{nishizaki2000}. 
Assuming that lattice strain or inhomogeneity leads
 to some smearing of the ideal first order phase transition
this small entropy change could easily be masked by the experimental noise.

We should note that in the present calculations we have included the influence 
of the magnetic field 
in the Zeeman energies only and neglected the orbital effects altogether.
This means that the vortex contribution to the condensation 
energy of both the (a) and (b) phases is assumed to be the same. 
There is no  reason to expect a significantly different vortex lattice response in 
the different phases, since the only changes are in the direction of
${\bf d}({\bf k})$ on the Fermi surface and not in its magnitude $|{\bf d}({\bf k})|$.

\section{Constraints on model parameters}

As mentioned above, the  experiments \cite{murakawa2004,xia2006} 
indicate that the region where the 
chiral $(a)$ state is stable is restricted to
the low field part of the $H-T$ plane below a critical field which must be in the range of
about $B_t\approx 10-20$ mT. 
This places a number of constraints on the possible parameter values which we
can use in our model Hamiltonian.

Using our model Hamiltonian, Eq.~\ref{hubbard},  we can ask what are the possible
values of the spin-dependent pairing interaction needed to qualitatively describe both
experiments.  In our model this is the ratio  
$U'/U \equiv U^{\uparrow\downarrow}/U^{\uparrow\uparrow}$
for opposite spin pairing compared to equal spin pairing. The other model parameters 
are  fixed beforehand, including the known band structure and the  
values of interaction parameters $U_{\parallel}$ and $U_{\perp}$ which were previously fitted
to other experiments \cite{annett2002}.  
For simplicity we have taken the spin-orbit coupling to have a fixed value, $\lambda =-0.02t$,
where $t=80meV$ is the $\gamma$ band in-plane nearest neighbor hopping integral 
in our tight binding band structure fit. 

\begin{figure}[tbp]
\epsfig{file=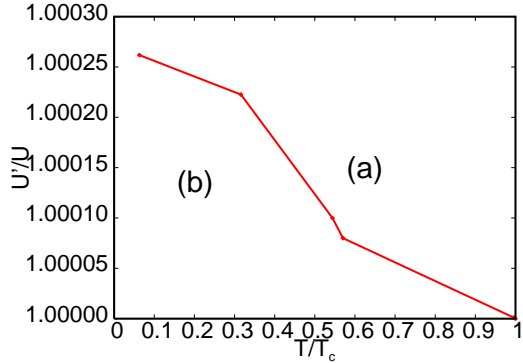,width=6.5cm,angle=-0}


\caption{(Color online) Temperature dependence of the spin dependent interaction 
anisotropy ratio $U'/U$  
necessary for $\lambda=-0.02t$ in the presence of external 
c-axis field ($H\parallel c$) to stabilize the chiral phase. 
The symbols $(a)$ and $(b)$ denote the stability regions of 
corresponding phases. 
\label{fig_upr}}
\end{figure}

With all other parameters fixed,  we then calculate the minimum spin dependent
interaction enhancement $U'/U$ required to stabilize the
chiral state in magnetic fields  of up to $B=20$ mT. 
It turns out that this minimal value of $U'/U$ changes with temperature, as shown in 
Fig.~\ref{fig_upr}.

\begin{figure}[tbp]
\epsfig{file=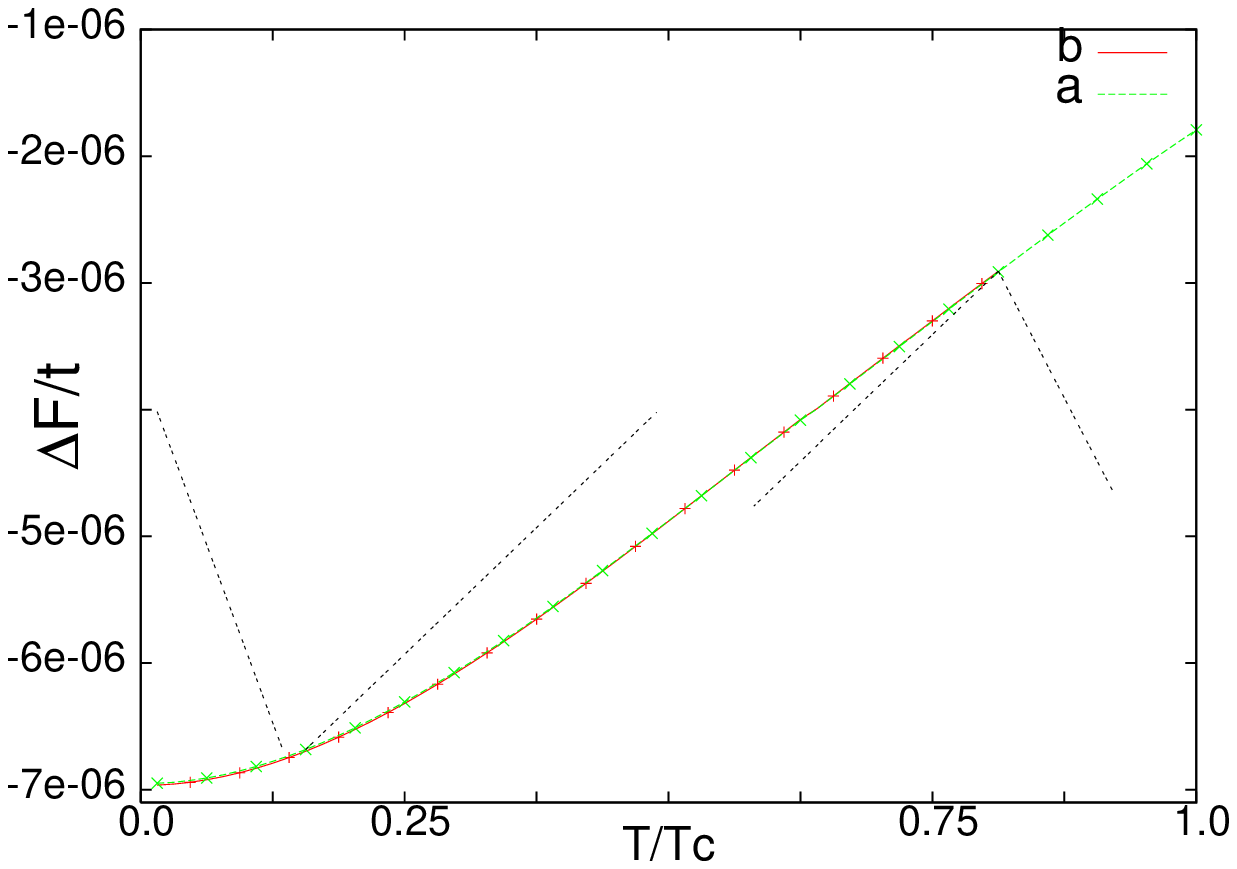,width=6.5cm,angle=-0}

\vspace{-2.05cm}
\hspace{3.1cm}
\epsfig{file=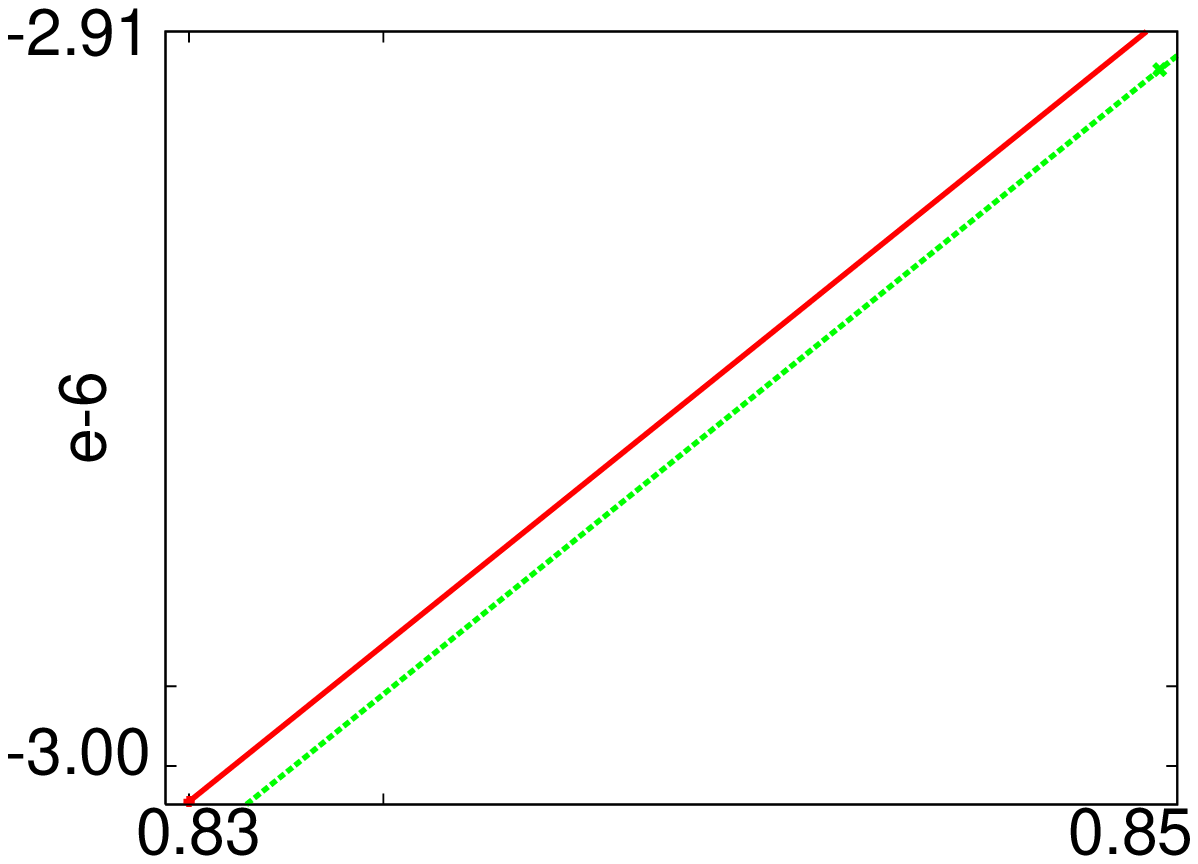,width=2.2cm,angle=-0}
\vspace{-4.0cm}

\hspace{-2.25cm}
\epsfig{file=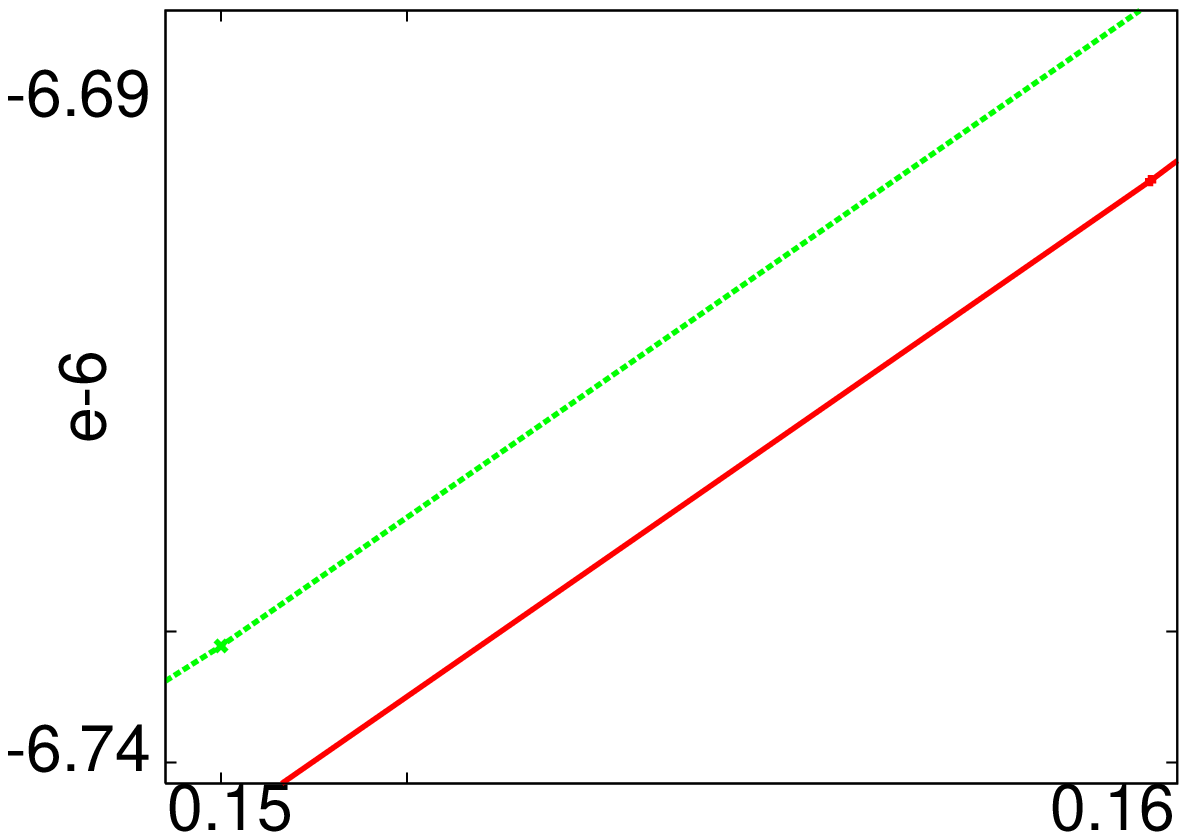,width=3.0cm,angle=-0}
\vspace{2.0cm}
\caption{(Color online) The temperature dependence of the condensation free 
energies for states 'a' and 'b' in a c-axis magnetic field $B=30$ mT 
for $\lambda=-0.02t$ and $U'=1.0001U$. The insets show the data on the
expanded scale for temperature well below  (upper left) 
and above crossing point (lower right).
\label{fig_fab}}
\vspace*{-0.5cm}
\end{figure}

The temperature dependence of $U'/U$ shown in Fig.~\ref{fig_upr} arises because the 
condensation energies calculated for a given 
value of $U'/U$ for $(a)$ and $(b)$
 states cross as function of temperature as shown in Fig.~\ref{fig_fab}.
For a small value of $U'/U$ for which the $(b)$ state 
will be  stable at low temperatures, we find that this state is destabilized 
relative to the chiral state closer to $T_c$.

Let us stress that the model we are using is, in principle at least, valid for 
arbitrarily large spin-orbit coupling. However, 
the parametrization of the Fermi surface which we have used ceases to be valid for
$|\lambda|>0.1t$.  Therefore in order 
not to refit the normal state band structure for every value of $\lambda$ we
have limited the calculations to small values of $\lambda$. 
In fact most of the calculations presented in this work
have been done  for $\lambda=-0.02t$. 
This restricted choice does not limit the validity of our conclusions,  because for every value 
of the spin-orbit coupling parameter, $\lambda$, it is possible to find a
corresponding value of  $U'/U$ which will stabilize the chiral state at $T=0K$
as discussed earlier ({\it c.f.} Fig. \ref{fig4}). Close enough to this
boundary of stability a c-axis magnetic field 
will rotate the d-vector
and stabilize  one of the other four states \cite{sigrist2005} independently
of the sign or magnitude
of $\lambda$. This minimal value of $U'/U$   
depends not only  on $\lambda$ and $B$ but also on temperature, as shown in Fig.~\ref{fig_upr}. 
However, this uncertainty in the value of $U'/U$ precludes an unique theoretical determination of 
the phase diagram on the $H-T$ plane without additional
assumptions. 

Assuming a constant ({\it i.e.} temperature independent) 
value of $U'/U$ which leads to stabilization of the
chiral phase at the lowest temperatures would lead to
a H-T phase diagram with the chiral 
phase occupying  most of it.  The non-chiral state
would be limited to low temperature-high field corner.  
Other assumptions about the temperature dependence of $U'/U$ could 
produce different topology phase diagrams.
Rather than try to explore all these possibilities we
simply assume that the boundary between the stable 
phases at low c-axis fields is given
by a line at approximately $B=B_t=20$ mT (shown in Fig.~\ref{fig2}) as  
dictated by experiment.  From this assumption we have then calculated the
corresponding temperature dependence of the minimal anisotropy $U'/U$ which is necessary 
to stabilize the a-phase along the
line (as shown in Fig.~\ref{fig_upr}). We find this approach more direct than 
trying to determine the expected
temperature dependence of $U'/U$ {\it a priori} from a spin fluctuation
feedback mechanisms \cite{goryo2002} 
stabilizing chiral phase.

We note that our effective pairing Hubbard model 
does not really include the full self-consistent effects of spin 
fluctuation feedback \cite{goryo2002}, which could make the effective pairing
interactions $U$ and/or $U'$ temperature dependent. However, the predicted
phase transition places severe constraints on the superconducting
pairing mechanism  when one  takes into account a sizable spin-orbit
coupling and the detrimental effect of the c-axis B-field on
the chiral phase.

\section{Summary and Conclusions}

To summarize, we have explored the role played by spin-orbit coupling in 
determining the superconducting states of strontium ruthenate in the presence
 of c-axis and ab-plane oriented magnetic fields.  We showed that the 
d-vector rotation can provide a consistent interpretation of  both the
 NQR and Kerr effect data, provided the spin-orbit coupling constant
 $\lambda$ is small enough. Unfortunately,  LDA estimates of $\lambda$ 
 are significantly larger than the values we have assumed in the calculations
 presented here\cite{pavarini2006}.  A large value of spin-orbit coupling has 
previously been cited as evidence against the d-vector rotation picture
\cite{zutic2005,kaur2005,pavarini2006}.  Despite these objections, it is 
clear from our results that even for  large values of $\lambda$ the transition 
field $B_t$ for d-vector rotation, in Fig.~\ref{fig2},
can lie within experimental constraints \cite{murakawa2004,xia2006}, but only 
if the pairing interaction spin anisotropy $U'/U$ is such as to
make the free energy difference between (a) and (b) phases finely balanced so
 that a small field of order $20$ mT
is sufficient to cause the d-vector rotation.  We have shown that for any value of $\lambda$
there is corresponding anisotropy $U'/U$ such that this balance can be achieved.
It is possible that such balancing is a consequence of a spin-fluctuation feedback 
mechanism \cite{goryo2002},
or emerges directly from the full microscopic spin fluctuation theory.
 But in the absence of a detailed theory of this effect
further progress in the field must await experimental measurement 
of the effective spin-orbit parameter
for the quasiparticles at the Fermi surface.\\

Acknowledgements: The authors acknowledge fruitful discussions 
 with Prof. Y. Maeno and Prof. I. Mazin. This work has been partially supported 
by the Ministry of Science and Higher Education grant No. N202 1878 33.


\end{document}